\documentclass[aps,superscriptaddress,nofootinbib,showpacs,eqsecnum,preprint,tightenlines]{revtex4}
\usepackage{hyperref}
\usepackage{epsfig,rotating}
\usepackage{amsmath,amssymb}
\usepackage{dsfont}
\usepackage{bbm}
\usepackage{slashed}
\numberwithin{equation}{section}

\newcommand{\be}{\begin{equation}}
\newcommand{\ee}{\end{equation}}
\newcommand{\bea}{\begin{eqnarray}}
\newcommand{\eea}{\end{eqnarray}}

\newcommand{\vp}{\vec{p}}

\newcommand{\vq}{\vec{q}}

\newcommand{\vk}{\vec{k}}

\newcommand{\m}{\mathbf{M}}

\newcommand{\Left}{{\mathds{L}}}
\newcommand{\Right}{{\mathds{R}}}

\begin{document}

\title{Charged lepton   mixing via heavy sterile neutrinos. }

\author{Louis Lello}
\email{lal81@pitt.edu}

\author{Daniel Boyanovsky}
\email{boyan@pitt.edu} \affiliation{Department of Physics and
Astronomy, University of Pittsburgh, Pittsburgh, PA 15260}

\date{\today}

\begin{abstract}
Pseudoscalar meson decay leads to an entangled state of charged leptons ($\mu,e$) and massive neutrinos. Tracing out  the neutrino degrees of freedom leads to a reduced density matrix for the charged leptons whose off-diagonal elements reveal \emph{charged lepton oscillations}. Although these decohere on unobservably small time scales $ \lesssim 10^{-23} s $ they indicate charged lepton \emph{mixing} as a result of common intermediate states. The charged lepton self energy up to one loop features flavor off-diagonal terms responsible for charged lepton mixing: a dominant ``short distance'' contribution with $W$ bosons and massive neutrinos in the intermediate state, and a subdominant ``large distance'' contribution with pseudoscalar mesons and massive neutrinos in the intermediate state.   Mixing angle(s) are GIM suppressed, and are   \emph{momentum and chirality dependent}. The difference of negative chirality mixing angles near the muon and electron mass shells is $\theta_L(M^2_\mu) -\theta_L(M^2_e)\propto G_F \sum U_{\mu j} m^2_j U^*_{j e}$ with $m_j$ the mass of the neutrino in the intermediate state.  Recent results from TRIUMF, suggest an upper bound $\theta_L(p^2\simeq M^2_\mu)-\theta_L(p^2 \simeq M^2_e) < 10^{-14}\,\Big(M_S/\mathrm{100}\,MeV\Big)^2$  for one generation of a heavy sterile neutrino with mass $M_S$. We obtain the wavefunctions for the propagating modes, and discuss the relation between the lepton flavor violating process $\mu \rightarrow e\gamma$ and charged lepton mixing, highlighting that a measurement of such process implies a mixed propagator $\mu, e$. Furthermore writing flavor diagonal vertices in terms of mass eigenstates yields novel interactions  suggesting further contributions to lepton flavor violating process as a consequence of   momentum and chirality dependent mixing angles.
\end{abstract}


\maketitle

\section{Introduction}

Neutrino masses, mixing and oscillations are the clearest evidence yet of physics beyond the standard model \cite{book1,revbilenky,book2,book3}.  Oscillations among three ``active'' neutrinos with $\delta m^2 = 10^{-4}-10^{-3}\,\mathrm{eV}^2$ for atmospheric and solar oscillations respectively have been firmly confirmed experimentally (see the reviews\cite{grimuslec}-\cite{lbe}).

However, several experimental hints have been accumulating that cannot be interpreted as mixing and oscillations among three ``active'' neutrinos with $\delta m^2 \simeq 10^{-4}-10^{-3}$. Early results from
 the LSND experiment\cite{lsnd} have recently been confirmed by   MiniBooNE running in antineutrino mode\cite{miniboone} both suggesting the possibility of new ``sterile'' neutrinos with $\delta m^2 \sim \mathrm{eV}^2$. The latest report from the MiniBooNE collaboration\cite{minilast} on the combined $\nu_\mu\rightarrow \nu_e$ and $\overline{\nu}_\mu\rightarrow \overline{\nu}_e$  \emph{appearance}  data is consistent with neutrino oscillations with $0.01 < \Delta m^2 < 1.0\,\mathrm{eV}^2$. This is consistent with the evidence from LSND antineutrino oscillations\cite{lsnd}, which bolsters the case for the existence of sterile neutrinos; however, combined MiniBooNE/SciBooNE analysis\cite{minisci} of the $\overline{\nu}_\mu$ \emph{disappearance} data are consistent with \emph{no short baseline disappearance} of $\overline{\nu}_\mu$. Recently, a re-examination of the antineutrino flux\cite{flux1} in anticipation of the Double Chooz reactor experiment resulted in a small increase in the flux of about $3.5\%$  for reactor experiments leading to a larger deficit of $5.7\%$ suggesting a \emph{reactor anomaly}\cite{reactor}. If this deficit is the result of neutrino mixing and oscillation  with baselines $L\lesssim 10-100\,\mathrm{m}$, it requires the existence of at least one sterile neutrino with $\delta m^2 \gtrsim 1.5 \,\mathrm{eV}^2$ and mixing amplitude $\sin^2(2\theta) \simeq 0.115$\cite{reactor}. Taken together these results may be explained by models that incorporate one or more sterile neutrinos that mix with the active ones\cite{sorel}-\cite{giuntishort} including perhaps non-standard interactions\cite{akshwet}; although, there is some tension in the sterile neutrino interpretation of short-baseline anomalies\cite{tension}. A  comprehensive review of short baseline oscillation experiments summarizes their interpretation in terms of one or more generations of sterile neutrinos\cite{conrad}.

  Recently it has been pointed out that the presence of sterile neutrinos may
 induce a modification of the recently measured angle $\theta_{13}$ \cite{wagner,gouvea}.

 Hints for the existence of sterile neutrinos also emerge  from cosmology. The   analysis of the cosmic microwave background anisotropies by WMAP\cite{wmap} suggests that the effective number of   neutrino species is $N_{eff} = 3.84\pm 0.40$ and $\sum(m_\nu)<0.44\,eV$, suggesting  the case for sterile neutrino(s) with $m \lesssim  \mathrm{eV}$, however the recent results from (SPT), (ACT)\cite{sptact}  and PLANCK\cite{planck} weaken the bounds considerably. Complementary cosmological data suggests that $N_{eff} >3$ at the $95\%$ confidence level\cite{cosmodata}, although  accommodating an $\mathrm{eV}$ sterile neutrino requires a reassessment of other cosmological parameters\cite{raffelt}. For recent reviews on ``light'' sterile neutrinos see ref.\cite{abarev}. Sterile neutrinos with masses in the $\sim \,\mathrm{keV}$ range \emph{may} also be suitable warm dark matter candidates\cite{dodelson}-\cite{dani} and appealing models of sterile neutrinos provide tantalizing mechanisms for baryogenesis\cite{drewes}.

 These   hints motivate several experimental proposals to search for sterile neutrinos (see the reviews in ref.\cite{abarev}). Various experimental searches have been proposed, such as Higgs decay and matter interactions of relic sterile neutrinos\cite{kuseexpt}, the end point of $\beta$-decay in $^{187}\mathrm{Re}$ with a value of $Q= 2.5 \,\mathrm{keV}$\cite{mare,hectormare},  electron capture decays of $^{163}Ho \rightarrow  ^{163}Dy$\cite{holmium}  and $^{8}Li$ production and decay\cite{litio}. More recently, the focus has turned on the possible new facilities at the ``intensity frontier'' such as project $X$ at Fermilab\cite{projectX}, alternative high intensity sources\cite{abarev,intensity} and recent proposals to study sterile-active oscillations with pion and kaon decay at rest (DAR)\cite{anderson,spitz} or muons from a storage ring\cite{storm} as well as the possibility of discrimination between heavy Dirac and Majorana sterile neutrinos via $|\Delta L|=2$ processes in high luminosity experiments\cite{dibmajo}, which is summarized in  a recent review\cite{conrad}. Although the recently reported analysis of the phase II data of the Mainz Neutrino Mass Experiment\cite{mainz} found no evidence for a fourth neutrino state tightening the limits on the mass and mixing of a fourth sterile species, the possibility of a \emph{heavy} sterile species is still actively explored\cite{gninenko,meloni}. More recently the PIENU collaboration at TRIUMF\cite{pienu} has reported an upper limit on the neutrino mixing matrix element $|U_{ei}|^2 \leq 10^{-8}$
 ($90\%\,C.L.$) in the neutrino mass region $60-129 \,\mathrm{MeV}/c^2$.

 In this article we focus on complementary consequences of sterile neutrinos in the form of \emph{charged lepton mixing phenomena}. The discussion of whether or not charged leptons \emph{oscillate} has been controversial\cite{pak}-\cite{field}, and more recently this question was addressed from the point of view of coherence\cite{akhos} highlighting that while oscillations are possible, they lead to rapid decoherence and no observable effects.
  Muon-antimuon oscillations via massive Majorana neutrinos have been studied in ref.\cite{love}, however,
   to the best or our knowledge the issue of charged lepton ($\mu-e$)  \emph{mixing} (we emphasize mixing over oscillations), has not yet received the same level of attention. Although in ref.\cite{boyhoas} charged lepton mixing and oscillations as a consequence of neutrino mixing was studied in early Universe cosmology at temperatures $m_\mu \ll T \ll M_W$ where it was argued that medium effects enhance charged lepton mixing,  the question of charged lepton \emph{mixing}    in vacuum and as a consequence of possible new generations of sterile neutrinos has not yet been discussed in the literature and is the main motivation of this article.

 Furthermore we discuss the relationship between the lepton flavor violating decay $\mu \rightarrow e\gamma$, and  charged lepton \emph{mixing} in terms of self-energies and propagators that mix $\mu$ and $e$. Charged lepton violation is the focus of current experimental   searches\cite{meg,meg2}, and a recent experimental proposal\cite{mu2e} to search for charged lepton flavor violation via the coherent conversion process $\mu- N \rightarrow  e- N$ at Fermilab.

 \vspace{2mm}

 \textbf{Goals:}
 In this article we study both charged lepton \emph{oscillations} and \emph{mixing} as a consequence of intermediate states of mixed massive neutrinos, and discuss the relationship between charged lepton mixing and charged lepton flavor violating processes.

  \begin{itemize}

\item \textbf{a) Oscillations:} In a recent article\cite{lello} (see also \cite{boya,desiternuestro}) we have provided a non-perturbative quantum field theoretical generalization of the Weisskopf-Wigner method to understand the correlated quantum state of charged leptons and neutrinos that consistently describes   pion/kaon decay in real time. Knowledge of this state allows us to obtain the reduced density matrix for charged leptons by tracing out the neutrino degrees of freedom. The off diagonal density matrix elements in the flavor basis contains all the information on charged lepton ($\mu, e$) \emph{coherence and oscillations}.

     \item \textbf{b) Mixing:} Charged lepton oscillations evidenced in the reduced density matrix are   a consequence of a  common set of intermediate states that couple to the charged leptons. We then study the charged current contribution to the one-loop  self-energy which couples charged leptons to an intermediate state of mixed massive neutrinos. The self-energy unambiguously determines the propagating states and  explicitly describe charged lepton \emph{mixing}. We obtain the mixed propagator, extracting the mixing angles and analyze the propagating modes and their wavefunctions. These results motivate us to address the relation between lepton flavor violating transitions such as $\mu \rightarrow e\gamma$ and charged lepton \emph{mixing}.

     \end{itemize}

 \vspace{2mm}

 \textbf{Brief summary of results:}

\begin{itemize}
\item \textbf{a:)}   The   quantum state of charged leptons and neutrinos from (light) pseudoscalar decay is a correlated entangled state from which we construct the corresponding (pure state) density matrix. Under the condition that neutrinos are not observed, we trace over their degrees of freedom leading to a reduced density matrix for the charged leptons. Because we focus solely on decay of $ \pi, K $ these are $\mu, e $. Integrating out the unobserved neutrinos leads to a reduced density matrix that is  off-diagonal   in the flavor basis. The off-diagonal   matrix elements describe charged lepton mixing and  exhibit   oscillations with typical frequency  $ E_{\mu}(k)-E_e(k) \gtrsim \mathcal{O}(m_{\mu} - m_e) \sim m_{\mu} \sim 1.6\times 10^{23} s^{-1}$ which  are unobservable over any experimentally relevant time scale and lead to rapid decoherence. This conclusion agrees with a similar observation   in ref.\cite{akhos}. While these fast oscillations lead to decoherence over microscopic time scales, we recognize that the \emph{origin} of these oscillations are a common set of intermediate states akin to neutral meson oscillations.

\item  Recognizing that the origin of oscillations are intermediate states that are common to both charged leptons we obtain the self-energy contributions and the full mixed propagator for the $\mu,e$ system. Mixing is a direct result of charged current interactions with intermediate neutrino \emph{mass eigenstates}.  As in the case of neutral meson mixing we identify ``short'' and ``long'' distance contributions to the flavor off-diagonal self-energies. The ``short'' distance contribution corresponds to the intermediate state of a $W^\pm$ and neutrino mass eigenstates and is dominant, whereas the ``long'' distance contribution is described by an intermediate state of $\pi,K$ and a neutrino mass eigenstate. We calculate explicitly the short distance and estimate the long distance contributions.  Unitarity of the neutrino mixing matrix entails a Glashow-Ilioupoulos-Maiani (GIM) type mechanism that suppresses   charged lepton mixing for light or nearly degenerate neutrinos, thus favoring heavy sterile neutrinos as intermediate states.

\item We obtain the flavor off-diagonal charged lepton propagator and analyze in detail the propagating modes.      $\mu-e$ mixing cannot be described solely in terms of a local  off-diagonal mass matrix but also off-diagonal \emph{kinetic terms} which are four-momentum dependent and contribute to off-shell processes. Mixing angles are GIM suppressed and both \emph{chirality and four momentum dependent}. The largest angle corresponds to the negative chirality component, the difference in mixing angles near the muon and electron mass shells is independent of the local renormalization counterterms and is given by $\theta_L(M^2_\mu)- \theta_L(M^2_e)\propto G_F \sum_j U_{\mu j} U^*_{j e} m^2_j$ where $m_j$ is the mass of the intermediate neutrino. Therefore charged lepton mixing is dominated by intermediate states with mixed heavy neutrinos. Assuming one generation of a heavy sterile neutrino with mass $M_S$ and extrapolating recent results from TRIUMF\cite{pienu} we obtain an upper bound   $\theta_L(M^2_\mu)- \theta_L(M^2_e) \lesssim 10^{-14}\,\Big(M_S/100\,\mathrm{MeV}\Big)^2$.
     We obtain the propagating eigenstates of charged leptons via two complementary methods: by direct diagonalization of the propagator and by field redefinitions followed by bi-unitary transformations, both results agree and yield momentum and chirality dependent mixing angles which are widely different on the respective mass shells.

\item The relationship between charged lepton \emph{mixing} and the lepton flavor violating decay $\mu \rightarrow e \gamma$ is discussed in terms of the mixed charged lepton self-energies and \emph{possible} observational effects in the form of further contributions to $\mu \rightarrow e \gamma$ are discussed. In particular we argue that writing the flavor lepton fields in terms of the propagating modes in flavor diagonal interaction vertices leads to novel interactions that depend on the difference of the mixing angles on the mass shells, this difference being independent of the choice of local renormalization counterterms.
\end{itemize}

\vspace{3mm}

\section{Reduced  Density Matrix: charged lepton oscillations}\label{sec:mudens}
In ref.\cite{lello} the quantum field theoretical Weisskopf-Wigner (non-perturbative) method has been implemented to obtain the quantum state resulting from the decay of a pseudoscalar meson $M$, (pion or kaon). It is found that such state is given by (see \cite{lello} for details and conventions),

\be
  |M^{-}_{\vp}(t))\rangle     =      e^{-iE_M(p)t} e^{-\Gamma_M(p)\frac{t}{2}} |M^{-}_{\vec{p}}(0)\rangle -  |\Psi_{l,\nu}(t)\rangle \label{Psidef} \ee where $|\Psi_{l,\nu}(t)\rangle$ is the \emph{entangled state} of charged leptons and neutrinos given by
   \be  |\Psi_{l,\nu}(t)\rangle    =     \sum_{j,\alpha, \vec{q},h,h'}\Bigg\{ U_{\alpha j}\, C_{\alpha j}(\vk,\vq,h,h';t)   |l^-_\alpha(h,\vk)\rangle\,|\overline{\nu}_j(h',-\vq)\rangle  \Bigg\}~~;~~\vk=\vp+\vq\, ,\label{finaresul} \ee  where
   \be  C_{\alpha,j}(\vk,\vq,h,h';t) =  \Pi_{\alpha j} \,\mathcal{M}_{\alpha j}(\vk,\vq,h,h')\,   \mathcal{F}_{\alpha j}[\vk,\vq;t] \,e^{-i(E_\alpha(k)+E_j(q))t}\,  \label{prodC}\ee with
\be \mathcal{F}_{\alpha j}[\vq,\vp,h,h';t] =
  \left[ \frac{1 - e^{-i(E^r_M(p) - E_{\alpha}(k) - E_j(q)- i \frac{\Gamma_M}{2}) t}}{E^r_M(p) - E_{\alpha}(k) -E_{j}(q) - i \frac{\Gamma_M}{2}} \right] \label{Fs}\ee and $\mathcal{M}_{\alpha,j}(\vk,\vq,h,h'),\Pi_{\alpha,j}(q,k)$ are the production matrix elements and phase space factors respectively,
  \be \mathcal{M}_{\alpha,j}(\vk,\vq,h,h')   =    F_M \,\overline{\mathcal{U}}_{\alpha,h}(\vk) \gamma^\mu \mathbbm{L} \mathcal{V}_{j,h'}(\vq)  p_\mu \label{prodmtxel}\ee \be  \Pi_{\alpha,j}(q,k)    =    \frac{1}{\sqrt{8VE_M(p)E_\alpha(k)E_j(q)}} \label{prodps}\ee

   In these expressions $F_M$ is the pion or kaon decay constant,  $\overline{\mathcal{U}};\mathcal{V}_{j,h'}(\vq)$ are the spinors corresponding to the charged lepton $\alpha$ and the neutrino mass eigenstate $j$ (for notation and details see ref.\cite{lello}). The \emph{leptonic} density matrix that describes the \emph{pure} quantum entangled state of neutrinos and charged leptons is given by
  \be \rho_{l,\nu}(t) = |\Psi_{l,\nu}(t)\rangle\langle\Psi_{l,\nu}(t)| \label{densmtx}\ee

If the neutrinos are \emph{not observed} their degrees of freedom must be traced out in the density matrix, the resulting density matrix is no longer a pure state,

\be
\rho^R_{l}(t)  = \mathrm{Tr}_{\nu} \rho_{l,\nu}(t) = \sum_{j,\alpha,\cdots}\sum_{j',\beta,\cdots} U_{\alpha j}\,U^*_{j \beta} C_{\alpha,j}C^*_{\beta,j}|l^-_\alpha \rangle \langle l^-_\beta | \langle \overline{\nu}_j| \overline{\nu}_{j'} \rangle  \,,  \label{redu1}\ee where $\langle \overline{\nu}_j| \overline{\nu}_{j'} \rangle = \delta_{jj'} $.

Considering only light pseudoscalar decay $\pi,K$, the only charged leptons available are $\mu,e$. For a fixed helicity $h$ and momentum $\vk$ of the charged leptons the reduced density matrix is given by
\bea
\rho^R_{l}(t)  & = &  \rho_{ee}(h,\vk,t) |e^-_{h,\vk}\rangle\ \langle e^-_{h,\vk}| + \rho_{\mu \mu}(h,\vk,t)|\mu^-_{h,\vk}\rangle\ \langle \mu^-_{h,\vk}| \label{rhored} \\ & + &  \rho_{e\mu}(h,\vk,t) |e^-_{h,\vk}\rangle\ \langle \mu^-_{h,\vk}|   + \rho_{\mu,e}(h,\vk,t) |\mu^-_{h,\vk}\rangle\ \langle e^-_{h,\vk}| \nonumber  \eea

The diagonal density matrix elements in the $\mu,e$ basis describe the \emph{population} of the produced charged leptons whereas the off-diagonal elements describe the \emph{coherences}. The diagonal matrix elements $\rho_{\alpha \alpha}~,~\alpha = \mu,e$ are given by

\be \rho^R_{\alpha \alpha}(t) = \sum_j |U_{\alpha,j}|^2 \, \mathrm{BR}_{M\rightarrow l_{\alpha} \nu_j} \big[1-e^{-\Gamma_M t}\big] ~~;~~ \alpha = \mu,e \label{diagels}\ee where $\mathrm{BR}$ are the branching ratios $\Gamma_{M\rightarrow l_{\alpha}~ \nu_j}/\Gamma_M$ and we have used some results obtained in ref.\cite{lello}. The off diagonal elements do not have a simple expression, however the most important aspect for the discussion is that these density matrix elements are of the form
\be \rho^R_{\mu e} = \sum_j U_{\mu j}\,U^*_{j e} C_{\mu,j}C^*_{e,j}~;~\rho^R_{e \mu} = (\rho^R_{\mu e})^* \,,\label{offdiag}\ee where the coefficients $C_{\alpha,j}$ are given by (\ref{prodC}-\ref{prodps}). These matrix elements  describe the   process $M \rightarrow \alpha  \nu_j$ followed by a ``recombination''-type process $ M  \nu_j \rightarrow \beta$ thereby suggesting the   intermediate state $M \nu_j$ common to both matrix elements.  For $\Gamma_M t \gg 1$  the reduced density matrix (for fixed $h,k$) in the charged lepton basis is of the form

\be
\rho^R = \left[ \begin{array}{cc} A_{ee} & A_{\mu e} e^{-i(E_{\mu}(k)-E_e(k))t} \\ A_{e \mu} e^{i(E_{\mu}(k) -E_e(k))t} & A_{\mu \mu} \end{array} \right]\,, \label{rhoredlt}
\ee

  This tells us that there will be $\mu \Leftrightarrow e$ oscillations. However,  these oscillations occur with large frequencies $ E_{\mu}(k)-E_e(k) \gtrsim \mathcal{O}(m_{\mu} - m_e) \sim m_{\mu} \sim 1.6\times 10^{23} s^{-1}$ and are unobservable over any experimentally relevant time scale. This conclusion agrees with a similar observation   in ref.\cite{akhos}.

  Although these oscillations average out over relevant time scales and are experimentally unobservable, an important issue is \emph{their origin}. The mixing between charged leptons arises from the fact that they share common \emph{intermediate states}, in the case studied above the common intermediate state corresponds to a pseudoscalar meson and a \emph{neutrino mass eigenstate}.

   Two aspects are important in the off diagonal terms in (\ref{offdiag}) whose long time limit defines $A_{\mu e}$:  a) from the expression (\ref{offdiag}) it follows that $A_{\mu e} \propto F^2_M $  and b) if all the neutrino states are degenerate the off diagonal terms vanish because the $C_{\alpha,j}$ would be the same for all $j$ and $\sum_j U_{\mu j}U^*_{j e} = 0$ by unitarity of the mixing matrix. This cancellation for massless or degenerate neutrinos is akin to the GIM mechanism.

   From this point of view the physical origin of the oscillations is found in \emph{mixing} of the charged leptons from the fact that they share common intermediate states. This is in fact similar to the oscillations and mixing through radiative corrections with common intermediate states in the $K_0 \overline{K}_0$ system. The obvious difference with this system is that, in absence of weak interactions, $K_0$ and $\overline{K}_0$ are degenerate and this degeneracy is lifted by the coupling to the (common) intermediate states, leading to oscillations on long time scales.

   The conclusion of this discussion   is that charged lepton oscillations are a result of their \emph{mixing} via a set of common intermediate states. The off-diagonal density matrix elements are of $\mathcal{O}(F^2_M)$, these are  the lowest order corrections in a perturbative expansion, therefore they do not reveal the full structure of the mixing phenomenon.

   If charged leptons mix via a common set of intermediate states, the correct propagating degrees of freedom are described by poles in the full charged lepton propagator which requires the self-energy correction. Such self-energy will reflect the \emph{mixing} through the intermediate states. Whereas oscillations average out the off-diagonal density matrix on short time scales, the main physical phenomenon of mixing is manifest in the   true propagating modes, namely the poles in the propagator which now becomes an off diagonal matrix in flavor space.

\section{Charged Lepton Mixing:}

\begin{figure}[h!]
\begin{center}
\includegraphics[keepaspectratio=true,width=2in,height=2in]{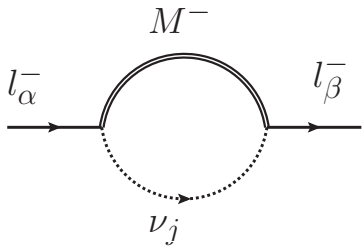}
\caption{Long distance contribution: intermediate state with $\nu_j$ and $M=\pi,K$. }
\label{fig:pionis}
\end{center}
\end{figure}

  We argued above that lepton mixing is a consequence of an intermediate meson/neutrino state  which couples to \emph{both} charged leptons. The intermediate meson state is a low energy or ``long distance'' representation of the coupling of charged leptons to quarks via charged current interactions and is akin to the mixing between $K_0 \overline{K}_0 $ via intermediate states with two and three pions.  This ``long distance'' (low energy) contribution to the charged lepton self energy is depicted in fig. (\ref{fig:pionis}).

\begin{figure}[h]
\begin{center}
\includegraphics[keepaspectratio=true,width=2in,height=2in]{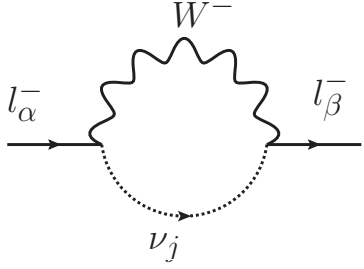}
\caption{Short distance contribution: intermediate state with $\nu_j$ and $W^-$.}
\label{fig:wis}
\end{center}
\end{figure}

 This  is   a low energy representation of   physical process in which a lepton couples to an intermediate W vector boson and a neutrino mass eigenstate, followed by the   decay of the (off-shell) W into   quark-antiquark pairs with the quantum numbers of the pseudoscalar mesons. Therefore we also expect a \emph{short distance} contribution in which the intermediate state corresponds simply to the exchange of a W boson and a neutrino mass eigenstate. This contribution to the charged lepton self-energy is depicted in fig. (\ref{fig:wis}).

We shall calculate both self-energy diagrams   to properly ascertain each  contribution to the full charged lepton propagator.

However, before carrying out the detailed calculations  we point out that there are also electromagnetic and neutral current contributions to the self-energies. However these are flavor \emph{diagonal}, thus while they will both contribute to the self-energies, only the charged current contributions  ( long and short distance) lead to  \emph{off diagonal} self energies which lead to charged lepton \emph{mixing}. Furthermore, both long and short distance self-energies are of the general form
\be \Sigma_{\alpha \beta} \propto \sum_j U_{\alpha j}\, S_j\, U^*_{j \beta } \label{segim}\ee where $S_j$ is the propagator of neutrino mass eigenstates, therefore unitarity of the mixing matrix $\sum_j U_{\alpha j}U^*_{j \beta} = \delta_{\alpha \beta }$ leads to a GIM (Glashow-Ilioupoulos-Maiani) type-suppression of the off-diagonal matrix elements: if all the neutrinos in the loop are degenerate, unitarity entails that there is no off-diagonal contribution to the self-energy, furthermore, this argument also suggests that the off-diagonal terms will be dominated by the most massive neutrino state.

Therefore charged lepton \emph{mixing} is a consequence of off-diagonal components of the self-energy matrix, which results in an off-diagonal \emph{propagator} for $e-\mu$ leptons as a consequence of common neutrino mass eigenstates in intermediate states. If the neutrinos are either massless or degenerate the unitarity of the mixing matrix leads to vanishing off-diagonal matrix elements and no mixing.

\subsection{Short distance contribution:}

We begin by computing the self energy contribution from W exchange depicted in fig. (\ref{fig:wis}). Throughout this calculation, we shall be working in the physical unitary gauge and in dimensional regularization.   Upon passing to the basis of mass eigenstates that define the neutrino propagators,
$\psi_{\nu_\alpha} = \sum_j U_{\alpha j} \psi_j$   the charged current contribution to the self energy \emph{matrix} is given by

\be
-i \Sigma_{\alpha \beta} = \left(\frac{-ig}{\sqrt{2}}\right)^2 \sum_j \int \frac{d^4k}{(2 \pi)^4} U_{\alpha j} \gamma_{\mu} \mathbbm{L} \left(\frac{i(\slashed{k}+m_j)}{k^2-m^2_j+i\epsilon}\right) U^*_{\beta j} \gamma_{\nu} \mathbbm{L} \left[\frac{-i\left(g^{\mu \nu} - \frac{k^{\mu} k^{\nu} }{M^2_W}\right)}{(p-k)^2 - M^2_W}   \right] \label{shdis}
\ee where $\mathbbm{L}, \mathbbm{R} = (1\mp \gamma^5)/2$ respectively.

This integral is calculated in dimensional regularization. We introduce a renormalization scale  $\kappa$ which we choose $\kappa = M_W$ thus renormalizing at the $W-pole$, and define
 \be \Delta_j = -p^2x(1-x) +m^2_j x +M^2_W (1-x) \,,\label{deltaj} \ee separating explicitly the divergent
  and finite parts in the $\overline{\mathrm{MS}}$ scheme we find
\be
 \Sigma_{\alpha \beta}(p) =    \slashed{p}\, \Left \Bigg[  \sum_j U_{\alpha j} U^*_{j \beta }     \int^1_0\,\left[ I^{d}_j(p^2;x)+I^{f}_j(p^2;x)\right]\, dx  \Bigg] \label{sigmashort}
\ee where
\be I^{d}_j(p^2;x) = -\frac{g^2\,M^{-\epsilon}_W}{2(4\pi)^2}(1-x)  \left[2 +\frac{3 \Delta_j}{M^2_W}+(1-x)^2\frac{ p^2}{M^2_W}\right]\left(\frac{2}{\epsilon} - \gamma + \ln{4\pi}\right) \label{Idiv} \ee
 \be I^{f}_j(p^2;x) = \frac{g^2 M^{-\epsilon}_W}{2(4\pi)^2}(1-x) \left[ \left(2+ \frac{3\Delta_j}{M^2_W} + (1-x)^2\frac{p^2}{M_W^2}\right)\ln\frac{\Delta_j}{M^2_W } + \frac{2x m^2_j}{M_W^2} \right]
 \,. \label{Ifinite}   \ee
 Unitarity of the neutrino mixing matrix in the form $\sum_j U_{\alpha j}U^*_{j \beta} = \delta_{\alpha \beta}$ leads to GIM-like cancellations in both the divergent and the finite parts for the off-diagonal components of the  self-energy matrix. Therefore for $\alpha \neq \beta$ we find
  \be \Sigma_{\alpha \beta}(p) =    \slashed{p}\, \Left\, \Big[z^d_{\alpha \beta}+z^f_{\alpha \beta}(p^2) \Big]\,, \label{lilz}\ee where
  \be z^d_{\alpha \beta}  =   -\frac{g^2\,M^{-\epsilon}_W}{64\pi^2}    \,  \sum_j U_{\alpha j} U^*_{j \beta } \frac{m^2_j}{M^2_W}   \left(\frac{2}{\epsilon} - \gamma + \ln{4\pi}\right)~~;~~\alpha \neq \beta \label{zdiv}\ee and
   \bea z^f_{\alpha \beta}(p^2) =  \frac{g^2\,M^{-\epsilon}_W}{32\pi^2} &&  \sum_j U_{\alpha j} U^*_{j \beta} \int^1_0 \Bigg\{ \Big[2+\frac{3 \Delta_j}{M^2_W} +(1-x)^2\frac{ p^2}{M^2_W}\Big] \ln\left(\frac{\Delta_j}{M^2_W} \right)+ \nonumber \\ && \frac{ 2 m^2_j}{ M^2_W}x \Bigg\}(1-x)\,dx ~~;~~\alpha \neq \beta \,.\label{zfini}\eea

\subsection{Long distance contribution:}

We now turn our attention to the    intermediate state described by the  exchange of a $\pi/K$ meson and a neutrino mass eigenstate. This is the state that suggested charged lepton mixing from the density matrix treatment in the previous section.  A difficulty arises in the calculation of the meson exchange because in order to properly describe the coupling between the meson and the charged lepton and neutrinos we would need the full off-shell form factor $F_M(q^2)$ which is a function of the loop momentum since the meson is   propagating off its mass shell in the intermediate state.   Clearly this is very  difficult to include in a reliable calculation,  therefore we restrict our study to an \emph{estimate} of this contribution obtained by simply using the on-shell value of the form factor, namely the meson decay constant $F_M$  in order to obtain an admittedly rough assessment of its order of magnitude.

 Under this approximation the  contribution to the self-energy matrix from this intermediate state is given by

\be
-i\Sigma^{M}_{\alpha \beta} = F^2_M \sum_j U^*_{\alpha j} U_{\beta j} \int \frac{d^4 k}{(2\pi)^4} (\slashed{p} - \slashed{k}) \mathbbm{L} \left(\frac{i(\slashed{k} +m_j)}{k^2 - m^2_j +i \epsilon} \right) (\slashed{p}-\slashed{k}) \mathbbm{L} \left( \frac{i}{(p-k)^2 -M^2_M + i\epsilon} \right)
\ee where $M_M$ is the meson mass. The width of the meson may be incorporated via a Breit-Wigner approximation $M_M \rightarrow M_M-i\Gamma_M/2$, however this will only yield a contribution which is higher order in $G_F$.

   The calculation is performed in dimensional regularization, choosing the renormalization scale $\kappa = M_W$ as for the short distance contribution,   introducing
\be \delta_j =   -p^2x(1-x) +m^2_j x +M^2_M (1-x) \,,\label{deltajmeson} \ee and separating the divergent and finite parts in the  $\overline{MS}$ scheme we find
\be
 \Sigma_{\alpha \beta}(p) =   \slashed{p}\, \Left \Bigg[  \sum_j U_{\alpha j} U^*_{j \beta }     \int^1_0 \left[J^d_j(p^2;x)+ J^f_j(p^2;x) \right] dx  \Bigg], \label{sigmalong}
\ee
where
\be  J^d_j(p^2;x) =  -\frac{ M^{2-\epsilon}_W  F^2_M}{(4\pi)^2} \Big(\delta_j(1+3x) -(1-x)x^2p^2\Big) \left(\frac{2}{\epsilon} - \gamma + \ln 4\pi \right)\,,\label{jotadiv}\ee
\be  J^f_j(p^2;x) =  \frac{ M^{2-\epsilon}_W  F^2_M}{(4\pi)^2} \Bigg[ 2x^2 \frac{\delta_j}{M^2_W} +\Bigg((1-x)x^2 \frac{p^2}{M^2_W}- (1+3x)\,\frac{\delta_j}{M^2_W}\Bigg)\,\ln \frac{\delta_j}{M^2_W}\Bigg]  \,. \label{jotafini}\ee
For the off-diagonal matrix elements, unitarity of the neutrino mixing matrix leads to GIM type cancellations as in the short distance case, therefore for $\alpha \neq \beta $ we find
  \be \Sigma_{\alpha \beta}(p) =   \slashed{p}\, \Left\, \big[\varsigma^d_{\alpha \beta}+\varsigma^f_{\alpha \beta}(p^2) \big]\,, \label{lilchiz}\ee where
  \be \varsigma^d_{\alpha \beta}  =  - 3\frac{ M^{2-\epsilon}_W  F^2_M}{32\pi^2}\, \sum_j U_{\alpha j} U^*_{j \beta } \frac{m^2_j}{M^2_W}\, \left(\frac{2}{\epsilon} - \gamma + \ln 4\pi \right)  ~~;~~\alpha\neq \beta \label{chidiv} \ee
\be\varsigma^f_{\alpha \beta}(p^2) =  \frac{ M^{2-\epsilon}_W  F^2_M}{16\pi^2}\,\sum_j U_{\alpha j} U^*_{j \beta } \, \int^1_0  \Bigg[ 2x^2 \frac{\delta_j}{M^2_W} +\Big((1-x)x^2 \frac{p^2}{M^2_W}- (1+3x)\,\frac{\delta_j}{M^2_W}\Big)\,\ln \frac{\delta_j}{M^2_W}\Bigg]  dx ~~;~~\alpha\neq \beta \label{chifinita} \ee

 However, with $F_M \propto G_F f_{\pi,K}$ and $f_{\pi,K} \sim \, 100 \, \mathrm{MeV}$ it follows that
\be F^2_M\,M^2_W \propto g^2 \Big(\frac{g f_{\pi,k}}{M_W}\Big)^2 \sim 10^{-8} g^2 \label{compa}\ee therefore the long distance contribution is negligible as compared to the short distance contribution and to leading order the off-diagonal components of the self energy are given by eqns. (\ref{lilz}- \ref{zfini}).

\vspace{1mm}

As noted previously   unitarity of the neutrino mixing matrix entails that the flavor off diagonal matrix elements of the self-energy vanish either for vanishing or degenerate neutrino masses. Obviously the contribution from light active-like neutrinos is strongly suppressed by the ratios $m^2_j/M^2_W$,  hence these off-diagonal matrix elements are dominated by the \emph{heaviest} species of sterile neutrinos.

\emph{Thus charged lepton mixing is enhanced by intermediate states with heavy sterile neutrinos}. This is one of the main results of this article.

\emph{If}  even the heaviest generation of sterile neutrinos feature masses $m_j \ll M_W$ and for $p^2 \ll M^2_W$ the following order of magnitude for the off-diagonal component $z_{\mu e}$ is obtained
\be z_{\mu e} \simeq \frac{G_F}{4\pi^2} \sum_j U_{\alpha j} U^*_{j \beta}\, m^2_j \,,\label{zetaapp}\ee as it will be seen below this estimate determines the mixing angles up to kinematic factors.

\section{Full propagator: mixing angles and propagating modes.}

\subsection{Full propagator and mixing angles:}

To treat $\mu,e$ mixing it is convenient to introduce a flavor doublet
\be \Psi = \left(
             \begin{array}{c}
               \psi_\mu \\
               \psi_e \\
             \end{array}
           \right)\,, \label{doublet} \ee
      The general structure of the self-energy is of the
form
\be \mathbf{\Sigma}(p)=    \Big[\mathbf{z}_L(p^2)\slashed{p}+\delta\mathbf{M}_L (p^2) \Big] \Left +    \Big[\mathbf{z}_R(p^2)\slashed{p}+\delta\mathbf{M}_R (p^2) \Big] \Right \,.    \label{gensigma}\ee
The neutral current interactions   contributes generally to the right and left components of the self-energy but are \emph{diagonal} in flavor and so are the  electromagnetic contributions. The   $V-A$ nature of the charged  current interactions is such that their contribution is only of the form $\mathbf{z}_L(p^2)  \slashed{p}\,\Left $ and is the \emph{only} contribution that yields flavor off-diagonal terms and are ultimately responsible for $\mu - e$ mixing. To cancel the poles in $\epsilon$ in the self-energy we allow  counterterms in the bare Lagrangian
\be \mathcal{L}_{ct} = \overline{\Psi} (\delta \mathbf{Z}_{ct}-1) \slashed{p} \Psi + \overline{\Psi}  \delta \m   \Psi + \mathrm{h.c.}\,. \label{counter} \ee

The full propagator $   \mathbf{S} $ now becomes a $2\times 2$ matrix which is  the solution of
\be \big[\slashed{p}\mathbf{1}+\slashed{p}(\delta \mathbf{Z}_{ct}-1) -\mathbf{\Sigma}(p)-\m\big]\mathbf{S} = \mathbf{1} \label{fullpropa}\ee where the boldfaced quantities are $2\times 2$ matrices and

\be \m =   \left(
                  \begin{array}{cc}
                    M_\mu & 0 \\
                   0 & M_e \\
                  \end{array}
                \right)
 \,. \label{massmatx}\ee
 In what follows we will assume that that $\m$ contains the renormalized masses and we will neglect finite momentum dependent contributions to $\m$ since  these will only generate higher order corrections to the mixing matrix as will become clear below.

  We will choose the counterterm $(\delta \mathbf{Z}_{ct}-1)$ in the $\overline{MS}$ scheme to cancel   the term $z^d_{\alpha \beta}$ in eqn. (\ref{lilz}). Therefore equation (\ref{fullpropa}) becomes
\be \Big[\slashed{p}\,{\mathbf{Z}}^{-1}_L \,\Left + \slashed{p}\, {\mathbf{Z}}^{-1}_R \,\Right - \m \Big]\mathbf{S} = \mathbf{1} \label{fullpropa2}\ee where
 \be   {\mathbf{Z}}^{-1}_{L,R} = \mathbf{1}-\mathbf{z}^f_{L,R}(p^2)\,.\label{capz}\ee  The leading contribution to the \emph{off-diagonal} matrix elements is given by the ``short-distance'' term eqn. (\ref{zfini}).

Multiplying on the left both sides of (\ref{fullpropa2}) by $\slashed{p} + \m {\mathbf{Z}}_R \,\Left + \m {\mathbf{Z}}_L\,\Right $ and writing the full propagator as
\be \mathbf{S}  =   \Right \mathbf{S}_{R}          + \Left \,\mathbf{S}_{L} \label{SRL}\ee where
\bea \mathbf{S}_R & = &   \mathbf{A}_{R}(p^2) \Big[\slashed{p}+\mathbf{B}_{R}(p^2)\Big]   \label{SR} \\  \mathbf{S}_L & = &  \mathbf{A}_{L}(p^2) \Big[\slashed{p}+\mathbf{B}_{L}(p^2)\Big]  \label{SL}\eea
 we find
\bea \Big(p^2 \,{\mathbf{Z}}^{-1}_R- \m {\mathbf{Z}}_L \m \Big) \mathbf{A}_R(p^2) & = & \mathbf{1} \label{ar}\\ \Big(p^2 \,{\mathbf{Z}}^{-1}_L- \m {\mathbf{Z}}_R \m \Big) \mathbf{A}_L(p^2) & = & \mathbf{1}
\label{al} \eea and the conditions
\be  \mathbf{B}_R(p^2)   =  \m\,{\mathbf{Z}}_L(p^2)\    ~~;~~
\mathbf{B}_L(p^2)  =   \m \, {\mathbf{Z}}_R(p^2)   \,.\label{condibs} \ee

 In what follows we will   neglect   CP violating phases in $U_{\alpha j}$ with the purpose of studying $\mu-e$ mixing in the simplest case. Under these approximations we find

\vspace{2mm}

\textbf{I):} The solution for $  \mathbf{A}_{R}(p^2)$ in eqn. (\ref{ar}) is obtained as follows. Consider the diagonalization of the inverse propagator

\be p^2 \,{\mathbf{Z}}^{-1}_R- \m {\mathbf{Z}}_L \m = \frac{1}{2}\Big[Q^R_\mu(p^2) +Q^R_e(p^2)\Big] \mathds{1} -\frac{\lambda_R(p^2)}{2} \, \left(
                          \begin{array}{cc}
                            \cos 2\,\theta_R(p^2) & \sin 2\,\theta_R(p^2) \\
                            \sin 2\,\theta_R(p^2) & -\cos 2\,\theta_R(p^2) \\
                          \end{array}
                        \right) \label{rightmtx} \ee where
 \be  Q^R_\alpha(p^2)   =   p^2 \big[{\mathbf{Z}}^{-1}_R \big]_{\alpha \alpha}-M^2_\mu \big[{\mathbf{Z}}_L  \big]_{\alpha \alpha} ~~;~~\alpha=\mu\,;\, e \label{QRalfa}\\
  \ee  and
  \be \lambda_R(p^2) =  \Bigg[ \Big(Q^R_{\mu}(p^2)-Q^R_e(p^2)\Big)^2 + 4\Big(M_\mu M_e \,{z}^f_{L,\mu e}(p^2) \Big)^2 \Bigg]^{\frac{1}{2}}\,. \label{lambdaR} \ee
 To leading order we find the mixing angle to be given by
\be \tan 2\,\theta_R(p^2) = \frac{2 M_{\mu} M_e \,z^f_{L,\mu e}(p^2)}{M^2_{\mu}-M^2_e} \,.   \label{Rangles}\ee The matrix above can be diagonalized by a unitary transformation
\be   \mathcal{U}[\theta] =   \left(
                          \begin{array}{cc}
                            \cos \theta & \sin \theta \\
                           - \sin \theta & \cos \theta \\
                          \end{array}
                        \right)   \label{rota}\ee

in terms of the mixing angle $\theta_R(p^2)$, namely
\be  \mathcal{U}[\theta_R(p^2)] \Big[p^2 \,{\mathbf{Z}}^{-1}_R- \m {\mathbf{Z}}_L \m \Big]   \mathcal{U}^{-1}[\theta_R(p^2)] =   \left(
                                                       \begin{array}{cc}
                                                         {Q^R_\mu(p^2)-\varrho_R(p^2)} & 0 \\
                                                         0 & {Q^R_e(p^2)+\varrho_R(p^2)} \\
                                                       \end{array}
                                                     \right) \label{Ritdiag} \ee
where to leading order
\be \varrho_R(p^2) = \frac{1}{2}(M^2_\mu - M^2_e) \tan^22\theta_R(p^2)\,, \label{xiR}\ee                                                   leading to the result
\be \mathbf{A}_R(p^2) = \mathcal{U}^{-1}[\theta_R(p^2)]\, \left(
                                                       \begin{array}{cc}
                                                         \frac{1}{Q^R_\mu(p^2)-\varrho_R(p^2)+i\epsilon} & 0 \\
                                                         0 & \frac{1}{Q^R_e(p^2)+\varrho_R(p^2)+i\epsilon} \\
                                                       \end{array}
                                                     \right)\,\mathcal{U}[\theta_R(p^2)]\,,\label{aRdiag1}\ee

which (to leading order)  simplifies to
\be \mathbf{A}_R(p^2) \simeq \mathcal{U}^{-1}[\theta_R(p^2)]\, \left(
                                                       \begin{array}{cc}
                                                         \frac{Z^R_{\mu \mu}(p^2)}{ p^2-M^2_{\mu}(p^2) -\varrho_R(p^2)+i\epsilon} & 0 \\
                                                         0 & \frac{Z^R_{ee}(p^2)}{ p^2-M^2_{e}(p^2) +\varrho_R(p^2)+i\epsilon} \\
                                                       \end{array}
                                                     \right)\,\mathcal{U}[\theta_R(p^2)]\,.\label{aRdiag}\ee
  In the above expressions $M^2_\mu(p^2),M^2_e(p^2)$ include the finite renormalization from the diagonal contributions of the self-energy matrix which have not been calculated here, furthermore the residues at the poles (wave-function renormalization) are also \emph{finite} since the (local) divergent contributions are canceled by the counterterm.

Therefore $\mathbf{S}_R$  can be written in the basis that diagonalizes the kinetic term
\be  \mathcal{U}[\theta_R(p^2)] \,\mathbf{S}_R \,\mathcal{U}^{-1}[\theta_R(p^2)]=   \left(
                                                       \begin{array}{cc}
                                                         \frac{Z^R_{\mu \mu}(p^2)\,\big[\slashed{p}+{b}^{R}_{\mu\mu}(p^2)\big]}{ p^2-M^2_{\mu}(p^2) -\varrho_R(p^2)+i\epsilon} &  \frac{Z^R_{\mu \mu}(p^2)\,{b}^{R}_{\mu e}(p^2)}{ p^2-M^2_{\mu}(p^2) -\varrho_R(p^2)+i\epsilon} \\
                                                        \frac{Z^R_{ee}(p^2)\,{b}^{R}_{e \mu}(p^2)}{ p^2-M^2_{e}(p^2) +\varrho_R(p^2)+i\epsilon} & \frac{Z^R_{ee}(p^2)\,\big[\slashed{p}+{b}^{R}_{e e}(p^2)\big]}{ p^2-M^2_{e}(p^2) +\varrho_R(p^2)+i\epsilon} \\
                                                       \end{array}
                                                     \right)\,   \label{Rpropa}  \,,                                      \ee

where
\be \mathbf{b}^R(p^2) = \mathcal{U}[\theta_R(p^2)] \, \m \mathbf{Z}_L(p^2) \, \mathcal{U}^{-1}[\theta_R(p^2)] \,. \label{lilbr}\ee

\vspace{2mm}

\textbf{II):} We proceed in the same manner for $\mathbf{A}_L(p^2)$, namely consider diagonalizing the inverse propagator

\be p^2 \,{\mathbf{Z}}^{-1}_L- \m {\mathbf{Z}}_R \m = \frac{1}{2}\Big[Q^L_\mu(p^2) +Q^L_e(p^2)\Big] \mathds{1} -\frac{\lambda_L(p^2)}{2} \, \left(
                          \begin{array}{cc}
                            \cos 2\,\theta_L(p^2) & \sin 2\,\theta_L(p^2) \\
                            \sin 2\,\theta_L(p^2) & -\cos 2\,\theta_L(p^2) \\
                          \end{array}
                        \right) \label{leftmtx} \ee where
\be  Q^L_\mu(p^2)   =   p^2 \big[{\mathbf{Z}}^{-1}_L \big]_{\alpha \alpha}-M^2_\mu \big[{\mathbf{Z}}_R  \big]_{\alpha \alpha} ~~;~~\alpha=\mu\,;\, e \label{QeL}\ee   and
  \be \lambda_L(p^2) =  \Bigg[ \Big(Q^L_{\mu}(p^2)-Q^L_e(p^2)\Big)^2 + 4\Big(p^2 \,{z}^f_{L,\mu e}(p^2) \Big)^2 \Bigg]^{\frac{1}{2}} \,.\label{lambdaL} \ee
 Again, to leading order we find the mixing angle to be given by
\be \tan 2\,\theta_L(p^2) = \frac{2p^2 \,z^f_{L,\mu e}(p^2)}{M^2_{\mu}-M^2_e}  \,.  \label{Langles}\ee  The matrix above can be diagonalized by the  unitary transformation (\ref{rota}) now in terms of the mixing angle $\theta_L(p^2)$, namely
\be \mathcal{U}[\theta_L(p^2)]\Big[ p^2 \,{\mathbf{Z}}^{-1}_L- \m {\mathbf{Z}}_R \m  \Big]\mathcal{U}^{-1}[\theta_L(p^2)] =  \left(
                                                       \begin{array}{cc}
                                                         {Q^L_\mu(p^2)-\varrho_L(p^2) } & 0 \\
                                                         0 &  {Q^L_e(p^2)+\varrho_L(p^2) } \\
                                                       \end{array}
                                                     \right)\,, \label{leftopdig}\ee
where to leading
\be \varrho_L(p^2) = \frac{1}{2}(M^2_\mu - M^2_e) \tan^22\theta_L(p^2)\, \label{xiL}\ee
leading to the result
\be \mathbf{A}_L(p^2) = \mathcal{U}^{-1}[\theta_L(p^2)]\, \left(
                                                       \begin{array}{cc}
                                                         \frac{1}{Q^L_\mu(p^2)-\varrho_L(p^2)+i\epsilon} & 0 \\
                                                         0 & \frac{1}{Q^L_e(p^2)+\varrho_L(p^2)+i\epsilon} \\
                                                       \end{array}
                                                     \right)\,\mathcal{U}[\theta_L(p^2)]\,,\label{aLdiag1}\ee

Neglecting the diagonal contributions to mass renormalization, but keeping the (finite) wave function renormalizations, the result (\ref{aLdiag1}) simplifies to
\be \mathbf{A}_L(p^2) \simeq \mathcal{U}^{-1}[\theta_L(p^2)]\, \left(
                                                       \begin{array}{cc}
                                                         \frac{Z^L_{\mu \mu}(p^2)}{ p^2-M^2_{\mu}(p^2) -\varrho_L(p^2)+i\epsilon} & 0 \\
                                                         0 & \frac{Z^L_{ee}(p^2)}{ p^2-M^2_{e}(p^2) +\varrho_L(p^2)+i\epsilon} \\
                                                       \end{array}
                                                     \right)\,\mathcal{U}[\theta_L(p^2)]\,,\label{aLdiag}\ee

Just as in the previous case, $M^2_\mu(p^2),M^2_e(p^2)$ include the finite contribution from mass terms in the self energy and the residues at the poles are also finite, the local, divergent contribution being canceled by the counterterm.

The   component  $\mathbf{S}_L$  can now  be written as
\be  \mathcal{U}[\theta_L(p^2)] \,\mathbf{S}_L \,\mathcal{U}^{-1}[\theta_L(p^2)] = \left(
                                                       \begin{array}{cc}
                                                         \frac{Z^L_{\mu \mu}(p^2)\,\big[\slashed{p}+{b}^{L}_{\mu\mu}(p^2)\big]}{ p^2-M^2_{\mu}(p^2) -\varrho_L(p^2)+i\epsilon} &  \frac{Z^L_{\mu \mu}(p^2)\,{b}^{L}_{\mu e}(p^2)}{ p^2-M^2_{\mu}(p^2) -\varrho_L(p^2)+i\epsilon} \\
                                                        \frac{Z^L_{ee}(p^2)\,{b}^{R}_{e \mu}(p^2)}{ p^2-M^2_{e}(p^2) +\varrho_L(p^2)+i\epsilon} & \frac{Z^L_{ee}(p^2)\,\big[\slashed{p}+{b}^{L}_{e e}(p^2)\big]}{ p^2-M^2_{e}(p^2) +\varrho_L(p^2)+i\epsilon} \\
                                                       \end{array}
                                                     \right)\,   \label{Lpropa}  \,,                                      \ee

where
\be \mathbf{b}^L(p^2) = \mathcal{U}[\theta_L(p^2)] \, \m \mathbf{Z}_R(p^2) \, \mathcal{U}^{-1}[\theta_L(p^2)] \,. \label{lilbl}\ee

An important aspect is that the mixing angles $\theta_R(p^2),\theta_L(p^2)$ not only are different for the $R,L$ components  a consequence of the $V-A$ nature of charged currents, but also that they feature very different momentum dependence,
\be \theta_R(p^2) \simeq \frac{M_e}{M_\mu} z^f_{L,\mu e}(p^2)~~;~~\theta_L(p^2) \simeq \frac{p^2}{M^2_\mu} z^f_{L,\mu e}(p^2) \label{tetas} \ee Near the muon mass shell $p^2 \simeq M^2_\mu$ it follows that $\theta_L \gg \theta_R$, for near the electron mass shell $p^2 \simeq M^2_e$ it follows that  $\theta_R \gg \theta_L$. Off-shell, for virtuality $p^2 \gg M^2_\mu$ mixing of the $L$ component becomes dominant.

In general the transformations that diagonalize  the kinetic terms $\slashed{p}$   for both the positive and negative chirality components \emph{do  not} diagonalize the mass terms. In the basis in which the kinetic terms are diagonal the pole-structure of the propagator is revealed and the propagating modes can be read-off. This basis, however,  does not diagonalize the mass term of the propagator and attempting to diagonalize the latter either via a unitary or a bi-unitary transformation will lead to an off diagonal matrix multiplying the kinetic term. A similar situation has been found in different contexts\cite{unsterile,sirlin,machet}.

\subsection{Propagating modes: the effective Dirac equation:}

The nature of the propagating modes is best illuminated by solving the effective Dirac equation for the flavor doublet, which corresponds to the zeroes of the inverse propagator, namely
\be \Big[\slashed{p}\,{\mathbf{Z}}^{-1}_L \,\Left + \slashed{p}\, {\mathbf{Z}}^{-1}_R \,\Right - \m \Big]\Psi(p) = 0 \,,\label{dirac}\ee with $\Psi$ a spinor doublet,
\be \Psi = \left(
                                     \begin{array}{c}
                                       \xi^R  \\
                                       \xi^L  \\
                                     \end{array} \right) ~~;~~ \xi^{R,L}  =  \left(
                                     \begin{array}{c}
                                       \xi^{R,L}_\mu  \\
                                       \xi^{R,L}_e  \\
                                     \end{array} \right)\,. \ee

It is convenient to work in the chiral representation and expand the  positive and negative chirality components in the helicity basis
\be \vec{\sigma}\cdot\,\frac{\vec{p}}{|\vec{p}|} v_{h}(\vec{p})  = h  v_{h}(\vec{p})~~;~~h= \pm 1 \label{heli}\ee in terms of which the spinor flavor doublet

\be \Psi(p) =    \sum_{h} v_h \otimes\, \left(
                                     \begin{array}{c}
                                       \xi^R_{h} \\
                                       \xi^L_{h} \\
                                     \end{array}
                                   \right)\,,\label{expaheli}\ee where $\xi^{R,L}_{h}$ are flavor doublets that obey the following equations
\bea {\mathbf{Z}}^{-1}_L (p_0+p \,h) \xi^L_{h} + \m     \xi^R_{h} & = & 0 \label{lefteq2}\\                              {\mathbf{Z}}^{-1}_R (p_0-p \,h) \xi^R_{h} + \m     \xi^L_{h} & = & 0 \label{righteq2}\,. \eea  The positive and negative energy and helicity components are given by ($p \equiv |\vec{p}|$)

\vspace{1mm}

\vspace{1mm}
\bea &&   \left(
                                     \begin{array}{c}
                                       \xi^R \\
                                       -\frac{\mathbf{Z}_L\,\m}{p_0+p} \,\xi^R \\
                                     \end{array}
                                   \right) ~~;~~ p_0>0 ; h=1 ~~;~~
 \left(
                                     \begin{array}{c}
                                       -\frac{\mathbf{Z}_R\,\m}{p_0+p}\,\xi^L \\
                                         \xi^L \\
                                     \end{array}
                                   \right) ~~;~~ p_0>0, h=-1 \label{solpos} \\ &&
                                   \left(
                                     \begin{array}{c}
                                        \frac{\mathbf{Z}_R\,\m}{|p_0|+p}\,\xi^L \\
                                         \xi^L\\
                                     \end{array}
                                   \right) ~~;~~ p_0<0 ; h=1 ~~~~;~~~~
 \left(
                                     \begin{array}{c}
                                       \xi^R \\
                                       \frac{\mathbf{Z}_L\,\m}{|p_0|+p} \,\xi^R \\
                                     \end{array}
                                   \right) ~~;~~ p_0<0, h=-1 \,.\label{solneg}   \eea

                                   The flavor doublets obey
\bea  \Big(p^2 \,{\mathbf{Z}}^{-1}_R- \m {\mathbf{Z}}_L \m \Big) \xi^R(p)  & = & 0 \label{xirite}\\ \Big(p^2 \,{\mathbf{Z}}^{-1}_L- \m {\mathbf{Z}}_R \m \Big)\xi^L(p)  & = & 0 \,,
\label{xilef}
 \eea using (\ref{Ritdiag}, \ref{leftopdig}) we find that the rotated doublets
 \be \mathcal{U}[\theta_R(p^2)] \left(
                                  \begin{array}{c}
                                   \xi^R_\mu(p) \\
                                    \xi^R_e(p) \\
                                  \end{array}
                                \right) = \left(
                                  \begin{array}{c}
                                   \varphi^R_1(p) \\
                                    \varphi^R_2(p) \\
                                  \end{array}
                                \right)
  ~~;~~ \mathcal{U}[\theta_L(p^2)] \left(
                                  \begin{array}{c}
                                   \xi^L_\mu(p) \\
                                    \xi^L_e(p) \\
                                  \end{array}
                                \right) = \left(
                                  \begin{array}{c}
                                   \varphi^L_1(p) \\
                                    \varphi^L_2(p) \\
                                  \end{array}
                                \right)\label{rotadoubles} \ee   obey the following equation
\be  \left(
                                                       \begin{array}{cc}
                                                         {Q^R_\mu(p^2)-\varrho_R(p^2)} & 0 \\
                                                         0 & {Q^R_e(p^2)+\varrho_R(p^2)} \\
                                                       \end{array}
                                                     \right)  \left(
                                  \begin{array}{c}
                                   \varphi^R_1(p) \\
                                    \varphi^R_2(p) \\
                                  \end{array}
                                \right) =0 \label{rightdireq} \ee

\be  \left(
                                                       \begin{array}{cc}
                                                         {Q^L_\mu(p^2)-\varrho_L(p^2)} & 0 \\
                                                         0 & {Q^L_e(p^2)+\varrho_L(p^2)} \\
                                                       \end{array}
                                                     \right)  \left(
                                  \begin{array}{c}
                                   \varphi^L_1(p) \\
                                    \varphi^L_2(p) \\
                                  \end{array}
                                \right) =0\,. \label{leftdireq} \ee     Neglecting perturbative renormalization of the $\mu,e$ masses, for $p^2 \simeq M^2_\mu$ the propagating modes correspond to  $\varphi^{R,L}_1 \neq 0~;~ \varphi^{R,L}_2  = 0 $   and the mixing angles for $R,L$ components are $\theta_{R,L}(M^2_\mu)$ respectively, with
\be \theta_R(M^2_\mu) \simeq \frac{M_e}{M_\mu}\,z^f_{L,\mu e}(M^2_\mu) ~~;~~ \theta_L(M^2_\mu) \simeq z^f_{L,\mu e}(M^2_\mu)\label{muangles} \ee  defining the $\mu$-like propagating modes
\be   \left(
                                  \begin{array}{c}
                                   \xi^L_\mu(p) \\
                                    \xi^L_e(p) \\
                                  \end{array}
                                \right) = \varphi^L_1(p) \left(
                                  \begin{array}{c}
                                    \cos\theta_L(M^2_\mu) \\
                                     \sin\theta_L(M^2_\mu) \\
                                  \end{array}
                                \right) ~~;~~  \left(
                                  \begin{array}{c}
                                   \xi^R_\mu(p) \\
                                    \xi^R_e(p) \\
                                  \end{array}
                                \right) = \varphi^R_1(p) \left(
                                  \begin{array}{c}
                                    \cos\theta_R(M^2_\mu) \\
                                     \sin\theta_R(M^2_\mu) \\
                                  \end{array}
                                \right)  \label{mulikemodes}\ee Similarly for $p^2 \simeq M^2_e$ the propagating modes near the electron mass shell correspond to $\varphi^{R,L}_2 \neq 0~;~ \varphi^{R,L}_1  = 0 $   and the mixing angles for $R,L$ components are $\theta_{R,L}(M^2_\mu)$ respectively, with
\be \theta_R(M^2_e) \simeq \frac{M_e}{M_\mu}\,z^f_{L,\mu e}(M^2_e) ~~;~~ \theta_L(M^2_e)  \simeq \frac{M^2_e}{M^2_\mu}~ z^f_{L,\mu e}(M^2_e)\label{eangles} \ee  defining the relation between the flavor doublets and the  propagating modes on the respective mass shells, namely
\be   \left(
                                  \begin{array}{c}
                                   \xi^L_\mu(p) \\
                                    \xi^L_e(p) \\
                                  \end{array}
                                \right) = \varphi^L_2(p) \left(
                                  \begin{array}{c}
                                     -\sin\theta_L(M^2_e)  \\
                                    \cos\theta_L(M^2_e) \\
                                  \end{array}
                                \right) ~~;~~  \left(
                                  \begin{array}{c}
                                   \xi^R_\mu(p) \\
                                    \xi^R_e(p) \\
                                  \end{array}
                                \right) = \varphi^R_2(p) \left(
                                  \begin{array}{c}
                                     -\sin\theta_R(M^2_e)  \\
                                    \cos\theta_R(M^2_e)  \\
                                  \end{array}
                                \right)  \label{elikemodes}\ee The expressions (\ref{mulikemodes},\ref{elikemodes}) combined with (\ref{solpos},\ref{solneg}) give a complete description of the propagating modes.

\subsection{{Alternative diagonalization procedure.}  } \label{subsec:alternative}
The quadratic part of the effective action in terms of the flavor doublet  (\ref{doublet}) and after renormalization is
\be \mathcal{L}_{eff} = \overline{\Psi}_R ~\slashed{p} ~ {\mathbf{Z}}^{-1}_R \,\Psi_R + \overline{\Psi}_L ~ \slashed{p} ~ {\mathbf{Z}}^{-1}_L \,\Psi_L - \overline{\Psi}_R \m \Psi_L - \overline{\Psi}_L \m \Psi_R \label{effaction}    \ee In this expression ${\mathbf{Z}}_{R,L}$ are finite because the renormalization counterterms cancelled the divergent parts. These finite wavefunction renormalization \emph{matrices} can be absorbed into a finite but four-momentum dependent renormalization of the Dirac fields, so that the kinetic terms are canonical, namely
\be  \overline{\Psi}_{R,L}(p)    =    \overline{\eta}_{R,L}(p)\, \sqrt{{\mathbf{Z}}_{R,L}(p)}~~;~~  {\Psi}_{R,L}(p) =  \sqrt{{\mathbf{Z}}_{R,L}(p)}~{\eta}_{R,L}(p)\,, \label{rightleftren}\ee leading to
\be \mathcal{L}_{eff} = \overline{\eta}_R \slashed{p}   \,\eta_R + \overline{\eta}_L \slashed{p}  \,\eta_L - \overline{\eta}_R \mathcal{M}(p) \eta_L - \overline{\eta}_L \,\mathcal{M}^\dagger(p)\, \eta_R \,, \label{effactionren}    \ee  we emphasize that because $\mathbf{Z}_L$ features off-diagonal terms, the above transformation is \emph{not only} a simple rescaling but also a \emph{mixing} between the $\mu,e$ fields.

The mass matrices
\be \mathcal{M}(p) = \sqrt{{\mathbf{Z}}_{R}(p)}~ \m ~\sqrt{{\mathbf{Z}}_{L}(p)} ~~;~~ \mathcal{M}^\dagger(p) = \sqrt{{\mathbf{Z}}_{L}(p)}~ \m ~\sqrt{{\mathbf{Z}}_{R}(p)} \label{masamtx}\ee
  feature off diagonal terms from $\mathbf{Z}_L$ and are \emph{momentum dependent}. They can be diagonalized by \emph{biunitary transformations}, namely introducing the \emph{unitary} matrices $\mathcal{V}_{R,L}$ as
 \be \eta_{R,L} =  \mathcal{V}_{R,L}\, \Phi_{R,L}~;~\overline{\eta}_{R,L} =    \overline{\Phi}_{R,L}\,\mathcal{V}^\dagger_{R,L} \label{Vmatx}\ee these matrices are momentum dependent and diagonalize the mass matrices,
 \be \mathcal{V}^\dagger_{R }\, \mathcal{M}\, \mathcal{V}_{L} = M_d ~~;~~\mathcal{V}^\dagger_{L }\, \mathcal{M}^\dagger\, \mathcal{V}_{R} = M_d \label{diagmtx}\ee where $M_d$ is a diagonal but momentum dependent ``mass'' matrix. It is straightforward to prove that
 \be \mathcal{V}^\dagger_{R }\, \mathcal{M}\,\mathcal{M}^\dagger\, \mathcal{V}_{R} = M^2_d = \mathcal{V}^\dagger_{L }\, \mathcal{M}^\dagger\,\mathcal{M} \, \mathcal{V}_{L}\,.\label{idtymtx}\ee
 Projecting the Dirac equation obtained from the effective action (\ref{effaction}) onto left and right handed components we find
 \be \big[p^2 - \mathcal{M}^\dagger\,\mathcal{M}\big]  \eta_L =0 ~~;~~  \big[p^2 - \mathcal{M} \,\mathcal{M}^\dagger \big]  \eta_R = 0 \, \label{DEs}\ee which are diagonalized by the unitary transformation (\ref{Vmatx}) with the property (\ref{idtymtx}). Obviously the position of the mass shells which are determined by the zeroes of the determinant of the operators in the brackets are the same as those obtained from the un-scaled Dirac equations (\ref{xirite},\ref{xilef}) a result that is straightforwardly confirmed.

 The $\mu$-like and $e$-like eigenvectores are
\be \Phi_R(p) = \varphi^R_1(p) \left(
                                  \begin{array}{c}
                                   1 \\
                                    0 \\
                                  \end{array}
                                \right) \,\mathrm{for}\, p^2 = M^2_\mu +\cdots ~~;~~ \Phi_R(p) = \varphi^R_2(p) \left(
                                  \begin{array}{c}
                                   0 \\
                                    1 \\
                                  \end{array}
                                \right) \,\mathrm{for}\, p^2 = M^2_e +\cdots  \label{righteigen}\ee      where the dots stand for the radiative corrections to the masses.  After straightforward algebra we find to leading order
 \be        \mathcal{V}_R[\delta_R] =   \left(
                          \begin{array}{cc}
                            \cos \delta_R & -\sin \delta_R \\
                             \sin \delta_R & \cos \delta_R \\
                          \end{array}
                        \right)  ~~;~~ \delta_R(p) \simeq \frac{  M_{\mu} M_e \,z^f_{L,\mu e}(p^2)}{M^2_{\mu}-M^2_e} \simeq  \frac{   M_e  }{M_ {\mu} }\,z^f_{L,\mu e}(p^2) \label{rotabiuR}\ee which is exactly the same as the rotation angle for the right handed component $\theta_R(p)$  given by eqn. (\ref{Rangles})  when evaluated on the   mass shells $p^2 \simeq M^2_\mu\,;\,p^2\simeq M^2_e$ respectively.

                        For the left handed component the $\mu$-like and $e$-like eigenvectores are
\be \Phi_L(p) = \varphi^L_1(p) \left(
                                  \begin{array}{c}
                                   1 \\
                                    0 \\
                                  \end{array}
                                \right) \,\mathrm{for}\, p^2 = M^2_\mu +\cdots ~~;~~ \Phi_R(p) = \varphi^L_2(p) \left(
                                  \begin{array}{c}
                                   0 \\
                                    1 \\
                                  \end{array}
                                \right) \,\mathrm{for}\, p^2 = M^2_e +\cdots  \label{lefteigen}\ee

and again to leading order we find
  \be        \mathcal{V}_L[\delta_L] =   \left(
                          \begin{array}{cc}
                            \cos \delta_L & -\sin \delta_L \\
                             \sin \delta_L & \cos \delta_L \\
                          \end{array}
                        \right)  ~~;~~ \delta_L(p) \simeq \frac{1}{2}\Bigg(\frac{  M^2_{\mu}+ M^2_e  }{M^2_{\mu}-M^2_e}\Bigg)\,z^f_{L,\mu e}(p^2)\,. \label{rotabiuL}\ee

We are now in position to reverse the re-scaling  and unitary transformation  to
 obtain the relation between the original $\mu,e$ fields and the fields that diagonalize the effective action, namely from (\ref{rightleftren}), and (\ref{Vmatx}) it follows that
 \be \Psi_{R,L} =   \left(
                                     \begin{array}{c}
                                       \xi^{R,L}_\mu  \\
                                       \xi^{R,L}_e  \\
                                     \end{array} \right) = \sqrt{{\mathbf{Z}}_{R,L}}~ \mathcal{V}_{R,L}\, \Phi_{R,L}\,. \ee Since ${\mathbf{Z}}_{R}$ is diagonal, we find to leading order

\be \left(
                                  \begin{array}{c}
                                   \xi^R_\mu(p) \\
                                    \xi^R_e(p) \\
                                  \end{array}
                                \right)_{p^2 \simeq M^2_\mu} \simeq  \varphi^R_1(p) \left(
                                  \begin{array}{c}
                                    1 \\
                                     \theta_R(M^2_\mu) \\
                                  \end{array}
                                \right) ~~;~~ \left(
                                  \begin{array}{c}
                                   \xi^R_\mu(p) \\
                                    \xi^R_e(p) \\
                                  \end{array}
                                \right)_{p^2 \simeq M^2_e} \simeq  \varphi^R_2(p) \left(
                                  \begin{array}{c}
                                   -\theta_R(M^2_e)\\
                                     1  \\
                                  \end{array}
                                \right)\,. \label{ritefina}\ee The matrix ${\mathbf{Z}}_{L}$ is off-diagonal so that    to leading order it follows that
\be \sqrt{{\mathbf{Z} }_{L}(p)} =  \left(
                                       \begin{array}{cc}
                                         1+\cdots  & \frac{1}{2}\,z^f_{L,\mu e}(p^2) \\
                                         \frac{1}{2}\,z^f_{L,\mu e}(p^2) & 1+\cdots \\
                                       \end{array}
                                     \right) \label{rutzl}
  \ee combining this result with (\ref{rotabiuL}) we find to leading order
\be \left(
                                  \begin{array}{c}
                                   \xi^L_\mu(p) \\
                                    \xi^L_e(p) \\
                                  \end{array}
                                \right)_{p^2 \simeq M^2_\mu} \simeq  \varphi^R_1(p) \left(
                                  \begin{array}{c}
                                    1 \\
                                     \theta_L(M^2_\mu) \\
                                  \end{array}
                                \right) ~~;~~ \left(
                                  \begin{array}{c}
                                   \xi^L_\mu(p) \\
                                    \xi^L_e(p) \\
                                  \end{array}
                                \right)_{p^2 \simeq M^2_e} \simeq  \varphi^L_2(p) \left(
                                  \begin{array}{c}
                                   -\theta_L(M^2_e)\\
                                     1  \\
                                  \end{array}
                                \right)\,  \label{leftyfina}\ee where $\theta_L(p^2)$ is given by eqn. (\ref{tetas}) with $p^2 \simeq M^2_\mu\,,\,M^2_e$ respectively.

 Thus we have confirmed that the alternative diagonalization procedure with rescaling the fields and diagonalizing the resulting mass matrices with bi-unitary transformations yield the same result as the direct procedure described in the previous sections, thereby establishing that the results obtained above are robust.

\section{Relation to lepton flavor violating processes:} 
\begin{figure}[h]
\begin{center}
\includegraphics[keepaspectratio=true,width=4.5in,height=4.5in]{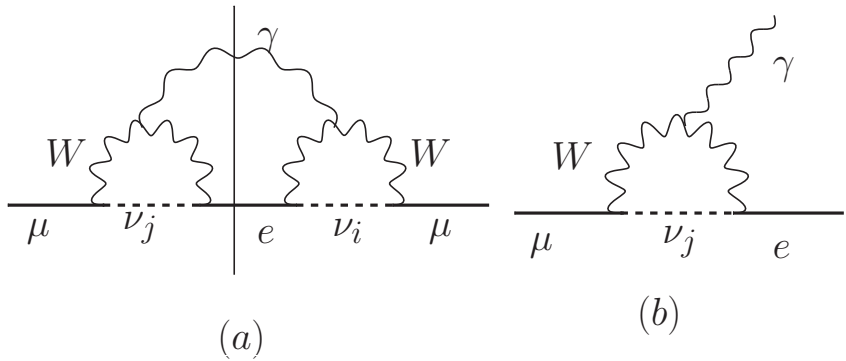}
\caption{Lepton flavor violation: fig. (a): three loop contribution to $\Sigma_{\mu \mu}$ the Cutkosky cut through the photon and electron intermediate state yields the imaginary part describing the flavor violating decay $\mu \rightarrow e\gamma$ of fig.(b). }
\label{fig:lfv}
\end{center}
\end{figure}

Charged lepton \emph{mixing} via intermediate states of charged vector bosons and neutrino mass eigenstates are \emph{directly} related to lepton flavor violating processes. An important process that is currently the focus of experimental searches\cite{meg,meg2} and a recent proposal\cite{mu2e} is the decay $\mu \rightarrow e\,\gamma$ which is mediated by neutrino mass eigenstates\cite{lfv1,lfv2,lfv3,lfv4} and the importance of heavy sterile neutrinos in this process has been highlighted in ref.\cite{ivan}. However, to the best of our knowledge the relationship between this process and a \emph{mixed} $\mu-e$ propagator has not yet been explored. Such relationship is best understood in terms of the \emph{three-loop} muon self-energy diagram in fig. (\ref{fig:lfv}-(a)), the Cutkosky cut along the intermediate state of the electron and photon yields the imaginary part of the muon propagator on its mass shell, and determines the decay rate $\mu \rightarrow e\gamma$, this is depicted in fig. (\ref{fig:lfv}-(b)).

\begin{figure}[h]
 \begin{center}
\includegraphics[keepaspectratio=true,width=3in,height=3in]{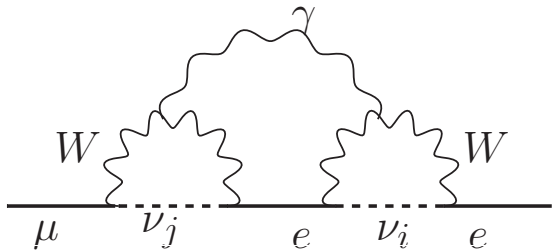}
\caption{Three loop contribution to $\Sigma_{\mu e}$  which is the off-diagonal counterpart of fig. (\ref{fig:lfv}(a).   }
\label{fig:lfvmue}
\end{center}
\end{figure}

However, the self-energy diagram (\ref{fig:lfv}-(a)) is only\emph{ one diagonal component} of the full $\mu-e$ self-energy, the corresponding three loop diagram for the off-diagonal component is shown in fig.
\ref{fig:lfvmue}.

 Because of the different external particles, a Cutkosky cut of this diagram through the photon and electron internal lines cannot be interpreted as a decay rate. However, this analysis clearly indicates the relationship between $\mu \rightarrow e\gamma$, a distinct indicator of lepton flavor violation, and charged lepton mixing in self-energy diagrams, both a direct consequence of neutrino mixing.

We note that whereas the branching ratio for $\mu \rightarrow e\gamma$ is $\propto G_F \alpha   |\sum_j U_{\mu j} U_{j e}|^2 m^2_j $ we find that the one-loop mixing angles are momentum dependent, different for different chiralities and the largest angle for on-shell states corresponds to the negative chirality muon-like combination, in which case the angle is of order $G_F  \sum_j U_{\mu j} U^*_{j e}  m^2_j $.

\vspace{1mm}

\begin{figure}[h]
\begin{center}
\includegraphics[keepaspectratio=true,width=3in,height=3in]{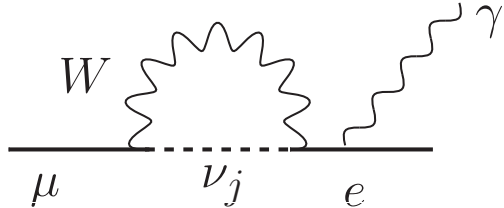}
\caption{ Further contribution to $\Sigma_{\mu e}$ from  $\mu -e$ mixing.   }
\label{fig:lfvnew}
\end{center}
\end{figure}

\textbf{\emph{Possible} other contributions:} The diagram in fig. (\ref{fig:lfv}-(b)) suggests that $\mu-e$ mixing \emph{may} lead to further contributions. Consider the flavor blind electromagnetic vertices of $\mu$ and $e$, \emph{if} the mixing angles were momentum independent, unitarity of the transformation  would entail a GIM cancellation between off-diagonal terms in the electromagnetic vertices, just as for neutral currents. However, muon-like and electron-like mass shells feature very different mixing angles which \emph{suggests} that off diagonal contributions arising from replacing the $\mu$ and $e$ fields in the electromagnetic vertices by the correct propagating states would \emph{not cancel out} because of different mixing angles. This can be seen from the relation between the propagating states and the $\mu,e$ states given by eqns. (\ref{mulikemodes},\ref{elikemodes}), writing
\be \psi_\mu = \cos\theta_1 \varphi_1 -\sin \theta_2 \varphi_2~~;~~ \psi_e =  \cos\theta_2 \varphi_2 +\sin \theta_1 \varphi_1 \label{relas}\ee respectively for positive and negative chirality components with the respective angles $\theta_{1L}=\theta_L(M^2_\mu);\theta_{2L}=\theta_L(M^2_e)$ etc., it follows that the electromagnetic vertices feature a mixed term of the form
\be \propto \overline{\varphi}_{2L} \gamma^\mu A_\mu \varphi_{1L} (\theta_{1L}-\theta_{2L})+ L \rightarrow R\,, \label{mixedem}\ee where the right handed angles are very different from the left handed counterparts.
If the mixing angle(s) were momentum independent $\theta_1=\theta_2$ and this term would vanish in a manner similar to the GIM mechanism. Furthermore the \emph{difference} $\theta_{1L}-\theta_{2L}$ is insensitive to the choice of the local renormalization counterterms.  Therefore mixing with \emph{momentum and chirality dependent mixing angles} suggests that the contribution to $\mu \rightarrow e \gamma$ from the vertex (\ref{mixedem})  depicted in fig. (\ref{fig:lfvnew}) becomes possible.

This contribution differs from that of fig. (\ref{fig:lfv}-(b)) in two major aspects: i) rather than an extra W propagator, it features an electron propagator in the intermediate state,  which would suggest a large enhancement with respect to the usual contribution, ii)  a very small mixing angle which suppresses the enhancement  from the electron propagator in the intermediate state. Thus a detail study of both effects and their impact is required for a firmer assessment.

 This argument, however, needs to be scrutinized further by analyzing the imaginary part of the propagators keeping both the diagonal electromagnetic contribution as well as the off diagonal charged current contribution. Upon the diagonalization of the propagator there is an interference between the diagonal and the off-diagonal terms, this can be gleaned from (\ref{rightmtx},\ref{leftmtx}).  The imaginary part of the propagator evaluated at the mass shell of the muon-like propagating mode, namely $p^2 \simeq M^2_\mu$ would yield the contribution to the diagram of fig. (\ref{fig:lfvnew}) from the interference between the diagonal  electromagnetic contribution which features an imaginary part for $p^2 > M^2_e$  and the off-diagonal charged current contribution. Since this is a contribution to the self energy of higher order than the ones considered here, a firmer assessment of this new contribution merits further study and will be reported elsewhere.

\section{Summary and discussion:}

We summarize and clarify some of the main results obtained above.

\begin{itemize}
\item{\textbf{Mixing:}} In this study mixing refers to the fact that the flavor eigenstates of charged leptons, $\mu;e$ are \emph{not the propagating states}. This is a consequence of self-energy corrections that are \emph{off-diagonal} in the flavor basis as a consequence of intermediate states with neutrino mass eigenstates that connect the flavor states. As is standard in quantum field theory, the propagating modes correspond to the poles of the full propagator, because of the off-diagonal self-energy contributions these propagators become an off diagonal matrix in flavor space, whose diagonalization yields the correct propagating modes. We offered two complementary methods to understand the mixing and diagonalization: (i) a direct diagonalization of the propagator matrix including the one-loop self energy which features the off-diagonal terms, (ii) a diagonalization of the effective action by first rescaling the fields to a canonical form followed by a bi-unitary transformation to diagonalize the mass terms. Both approaches yield the same result: mixing angles that depend on the corresponding mass shells and different for right and left-handed components, these are given by (\ref{muangles},\ref{eangles}) which are also obtained via the procedure of rescaling the fields to a canonical form, diagonalizing the mass matrices by a bi-unitary transformation and re-scaling back to find the relation between the original flavor eigenstates and the propagating eigenstates yielding the same mixing angles (see the discussion below eqns. (\ref{rotabiuR},\ref{leftyfina}).  As described with both methods, the transformations necessary to relate the flavor and propagating eigenstates are manifestly \emph{non-local} which is reflected on the different mixing angles on the different mass shells.

    The mixing angles are GIM suppressed favoring heavier neutrinos in the intermediate state and \emph{momentum and chirality dependent}. This means that off-shell processes necessarily mix charged leptons with \emph{virtuality and chirality dependent mixing angles}. For $p^2 \ll M^2_W$ and assuming that the heaviest sterile neutrinos feature masses $\ll M^2_W$, from eqn. (\ref{zetaapp})  we find the positive and negative chirality mixing angles for $\mu-e$ mixing
    \be \theta_R \simeq \frac{G_F}{4\pi^2}\,\frac{M_e}{M_\mu} \,\sum_j U_{\mu j} U^*_{j e}\, m^2_j  ~~;~~ \theta_L(p^2) \simeq  \frac{G_F}{4\pi^2} \frac{p^2}{M^2_\mu}\,\sum_j U_{\mu j} U^*_{j e}\, m^2_j  \label{mixan}\ee thus the mixing angles are dominated by the heaviest generation of neutrinos, and the \emph{difference} of mixing angles at the different mass shells is insensitive to the choice of local renormalization counterterms. In particular \emph{if} heavy sterile neutrinos do exist, these new degrees of freedom will yield the largest contribution to  charged lepton mixing.  Considering for example that there is only one generation of heavy sterile neutrinos with mass $M_S$, and \emph{assuming} that $U_{\mu i} \simeq U_{ei}$, the recent results from the PIENU collaboration at TRIUMF\cite{pienu} reporting an upper limit $|U_{ei}|^2 \leq 10^{-8}$
 ($90\%\,C.L.$) in the neutrino mass region $60-129 \,\mathrm{MeV}/c^2$ allows us to estimate an upper bound for  the negative chirality mixing angle near the $\mu$ mass shell,
 \be \theta_L(p^2\simeq M^2_\mu) - \theta_L(p^2\simeq M^2_e)\leq 10^{-14}\,\Big(\frac{M_S}{100\,\mathrm{MeV}}\Big)^2 \,.\label{upboundtetal}\ee

 \textbf{Oscillations}  are manifest in the off diagonal density matrix elements in the \emph{flavor basis}. These, however, average out on unobservable small time scales thus coherence (off-diagonal density matrix elements in the flavor basis) is suppressed by these rapid oscillations and is not experimentally relevant.

 \item{\textbf{Renormalization:}} The off-diagonal component of the self-energy (in the $\mu-e$ basis), features ultraviolet divergences, which are regularized in dimensional regularization consistently with the underlying gauge symmetry. The renormalization counterterm has been chosen in the $\overline{MS}$ scheme as is commonly done. The fact that the renormalized Lagrangian requires an off-diagonal counterterm is again a consequence of the fact that intermediate states with neutrino mass eigenstates mix the flavor fields $\mu-e$. However, the counterterm in the renormalized Lagrangian is \emph{local} and cannot completely remove the mixing between the flavor fields, this is manifest in the non-local and finite contribution to the off-diagonal self-energy given by eqn. (\ref{zfini}) for the short distance contribution and (\ref{jotafini}) for the long-distance contribution. These \emph{finite} contributions are momentum dependent and feature absorptive cuts  above the two particle threshold corresponding to the intermediate state of a charged vector boson and a neutrino mass eigenstate.

     The momentum dependence leads to the different mixing angles on the mass shells as discussed in detail in the previous section, and the  absorptive part gives rise to off-shell processes that involve the mixing of the flavor fields. In particular the \emph{difference} between the mixing angles at the two mass shells is independent of the local counterterm which is also obviously irrelevant for the absorptive part.

 \item{\textbf{Lepton flavor violation: }} The relationship between the off-diagonal self energy and lepton flavor violating processes becomes manifest by explicitly comparing the Feynman diagram in fig. (\ref{fig:wis})  for the self energy with $l^-_{\alpha} = \mu;l^-_{\beta} = e$ with that of the lowest order lepton flavor violating process $\mu \rightarrow e\gamma$  in fig. (\ref{fig:lfv} -(b)): neglecting the photon line, the intermediate state of $W-\nu_j$ is the \emph{same} as for the self-energy  (\ref{fig:wis}), namely: the mixing of flavors as a consequence of an off-diagonal self-energy in the $\mu-e$ basis has the same \emph{physical origin} as the lepton-flavor violating process $\mu \rightarrow e\gamma$. The direct relationship between the off-diagonal self-energy and
     $\mu \rightarrow e\gamma$ is shown explicitly in figs. (\ref{fig:lfv},\ref{fig:lfvmue} ).   Diagram (\ref{fig:lfv}-(a)) is the $\mu-\mu$ (diagonal) part of the self-energy, its Cutkosky cut across the W-line yields the imaginary part describing the process $\mu \rightarrow e\gamma$ in (\ref{fig:lfv}-(b)). The \emph{same} types of intermediate states yield the \emph{off-diagonal} $\mu-e$ contribution to the self-energy, displayed in fig.(\ref{fig:lfvmue} )  clearly indicating  that the \emph{physical origin} of the \emph{mixing} of $\mu-e$ flavor fields is the \emph{same} as the lepton flavor violating transitions $\mu \rightarrow e\gamma$.
     Upon writing the charged lepton fields in terms of the propagating modes in flavor diagonal vertices in the interaction Lagrangian, the momentum dependent field redefinition associated with the rescaling and bi-unitary transformation, namely the mixing, yields novel interaction vertices in terms of the propagating modes that depend on the \emph{difference} of the mixing angles at the different mass shells,  this difference is independent of the \emph{local} renormalization counterterm. A simple example is the electromagnetic vertex which is flavor diagonal, upon writing it in terms of the propagating modes $\varphi_{1,2}$ it describes an interaction between these in terms of the difference between the mixing angles at the mass shells, see eqn. (\ref{mixedem}) that leads to potentially new observable contributions such as that displayed in fig.(\ref{fig:lfvnew}) that merit further study.

\end{itemize}

\section{Conclusions and further questions}

In this article we studied charged lepton \emph{oscillations} and \emph{mixing}. The decay of pseudoscalar mesons leads to an entangled quantum state of neutrinos and charged leptons (we focused on $\pi,K$ decay leading to $\mu,e$). If the neutrinos are not observed, tracing over their degrees of freedom leads to a density matrix for the charged leptons whose off-diagonal elements in the flavor basis reveals \emph{charged lepton oscillations}. While these oscillations decohere on unobservably small time scales $\lesssim 10^{-23}\,s$, we recognize that they originate in a common set of intermediate states for the charged leptons. This realization motivated us to study the mixed $\mu-e$ \emph{self-energies} and we recognized that charged-current interactions lead to a dominant ``short distance'' contribution to $\mu-e$ mixing via W-exchange and an intermediate neutrino mass eigenstate, and a subdominant (by a large factor) ``long distance'' contribution to mixing via an intermediate state with a pseudoscalar meson and neutrino mass eigenstate. We include the leading contribution in the propagator matrix for the $\mu-e$ system focusing on the off-diagonal terms which imply $\mu-e$ mixing. We find that the mixing angles are \emph{chirality and momentum dependent}, the chirality dependence is a consequence of $V-A$ charge current interactions.  Diagonalizing the kinetic term and the mass matrix by bi-unitary transformations or alternatively diagonalizing the propagator, displays the poles which describe
muon-like and electron-like propagating modes (``mass eigenstates'') for which we find explicitly the wave functions,  but the mixing angles evaluated on the respective mass shells (and chiralities) are very different.  We find the positive and negative chirality momentum dependent mixing angles for $p^2 \ll M^2_W$ to be approximately given by
 \be \theta_R \simeq \frac{G_F}{4\pi^2}\frac{M_e}{M_\mu}\, \sum_j U_{\mu j} U^*_{j e}\, m^2_j  ~~;~~ \theta_L(p^2) \simeq  \frac{G_F}{4\pi^2} \frac{p^2}{M^2_\mu}\,\sum_j U_{\mu j} U^*_{j e}\, m^2_j  \label{mixanfin}\ee therefore dominated by the heaviest generation of sterile neutrinos. The \emph{difference} of mixing angles at the different mass shells is independent of local renormalization counterterms.  For one (dominant) generation of massive  sterile neutrinos with mass $M_S$, the recent results from the PIENU collaboration at TRIUMF\cite{pienu}, suggests
 \be \theta_L(p^2\simeq M^2_\mu)-\theta_L(p^2 \simeq M^2_e) \leq 10^{-14}\,\Big(\frac{M_S}{100\,\mathrm{MeV}}\Big)^2 \,. \ee

 Flavor diagonal interaction vertices feature novel interactions once written in terms of the fields associated with the propagating modes or mass eigenstates. In particular the electromagnetic vertex, yields an interaction between the muon-like and electron-like propagating modes which is another manifestation of lepton flavor violation. The (four) momentum dependence of the $\mu-e$ mixing angle \emph{may} be the source of novel off-shell effects whose potential observational manifestation merits further study. We expect to report on ongoing   study on these issues elsewhere.

 We discussed the relationship between the lepton flavor violating decay $\mu \rightarrow e\gamma$, the focus of current searches\cite{meg,meg2} and proposals\cite{mu2e}, and charged lepton mixing, pointing out that a positive measurement of the former confirms the latter. Furthermore, we advance the possibility of further contributions to $\mu \rightarrow e\gamma$ arising from the fact that the $\mu-e$ mixing angle is momentum dependent and differs substantially on the mass shells of the propagating modes voiding  a GIM mechanism for the electromagnetic
 vertices.

 Furthermore, in order to present the main arguments in the simplest case,  in this article we have not considered CP-violating phases in the mixing matrix elements $U_{\alpha j}$,   including these phases
merit further study since this aspect could indicate potentially   rich CP-violating phenomena from the charged lepton sector \emph{induced} by CP-violation from the neutrino sector which merits further and deeper study.

\acknowledgments We thank Adam Leibovich for stimulating conversations, Robert Shrock for illuminating correspondence.   The authors   acknowledge support from NSF through  grant PHY-1202227.

\end{document}